\documentclass[showpacs,twocolumn,prl,superscriptaddress]{revtex4-1}

\usepackage{slashed}
\usepackage{mathrsfs}
\usepackage{amsmath}
\usepackage{amssymb}
\usepackage{revsymb}
\usepackage{graphicx,accents}
\usepackage{graphicx,epstopdf}
\usepackage{mathrsfs}
\usepackage{bm}
\usepackage{psfrag}
\usepackage{hyperref}
\hypersetup{colorlinks=true,citecolor=Red,urlcolor=Red}
\usepackage{bbm}
\usepackage{bbold}
\usepackage{tabularx}
\usepackage{graphicx}
\usepackage[normalem]{ulem}
\usepackage[usenames,dvipsnames]{xcolor}

\usepackage[usenames,dvipsnames]{xcolor}
\usepackage{multirow,tabularx}
\usepackage{xcolor,pifont}
\newcommand*\colourcheck[1]{%
  \expandafter\newcommand\csname #1check\endcsname{\textcolor{#1}{\ding{52}}}%
}

\newcommand{\be}{\begin{equation}}
\newcommand{\ee}{\end{equation}}
\newcommand{\bea}{\begin{eqnarray}}
\newcommand{\eea}{\end{eqnarray}}

\newcommand{\trm}[1]{\textrm{#1}}

\newcommand{\LCperp}{{\scriptscriptstyle \perp}}
\newcommand{\LCm}{{\scriptscriptstyle -}} 
\newcommand{\LCp}{{\scriptscriptstyle +}}
\newcommand{\LCpm}{{\scriptscriptstyle \pm}}

\newcommand{\ud}{\mathrm{d}}

\newcommand{\vepsilon}{\varepsilon}

\definecolor{bk1}{RGB}{0,200,100}

\begin{document}
\title{Finite beaming effect on QED cascades}
\author{Suo Tang}
\email{tangsuo@ouc.edu.cn}
\affiliation{College of Physics and Optoelectronic Engineering, Ocean University of China, Qingdao, Shandong, 266100, China}


\begin{abstract}
The quantum electrodynamic (QED) theory predicts the photon emission and pair creation involved in QED cascades occur mainly in a forward cone with finite angular spread $\Delta\theta \sim 1/\gamma_{i}$ along the momenta of incoming particles. This finite beaming effect has been assumed to be negligible because of the particles' ultra-relativistic Lorentz factor $\gamma_{i}\gg1$ in laser-driven QED cascades. We develop an energy- and angularly resolved particle-tracking code, resolving both the energy spectra and the momentum profile of the outgoing particles in each QED event, which improves substantially the agreement between the simulation and exact QED results. We investigate QED cascades driven by two counter-propagating circularly polarized laser pulses, and show that the narrow beaming could be accumulated to effectively suppress the long-term growth of cascades, even though it can hardly affect the early formation of cascades. For QED cascades longer than $10$ laser cycles, the finite beaming effect could decrease the final pair yield, especially at ultrahigh intensities $\xi>600$, by more than $10\%$.
\end{abstract}
\maketitle
%
\section{Introduction}
The new generation of laser facilities, such as Vulcan~\cite{Vulcan10}, SULF~\cite{SULF}, and XCELS~\cite{XCELS2021}, are going to produce multi-$10$ PW laser pulses with focal strength over $10^{24}~\trm{W}/\trm{cm}^{2}$, which would open the avenue to investigate strong-field quantum electrodynamic (QED) effects in the fully nonperturbative regime~\cite{RMP1177,Fedotov:2022ely,RMP2022_045001} and reproduce in laboratories astrophysical phenomena in extreme environments~\cite{RMP78755}.
QED cascades occurring in astrophysical environment like a pulsar's magnetosphere~\cite{goldreich1969pulsar,wardle1998electron}, could be driven by ultrarelativistic laser fields initialized by few probe seeds, caused by avalanche events of hard photon emission and electron-positron pair creation~\cite{PRL2008200403,PRA042117,PoP5022640,Guo2024}.
The dynamics of QED cascades is complicated and characterized by a nonlinear interplay between the strong-field QED effects and classical plasma dynamics~\cite{Ehlotzky_2009,PRL080402,PRL035001}.

QED cascades have been widely studied in different field configurations via the Boltzmann-type kinetic equation~\cite{PoP3624481,PRA2013062110,Seipt2021} or particle-tracking simulations based on Monte-Carlo algorithm because of the stochastic nature of the QED events~\cite{PRA022105,PRE023210,PRE031201}.
One of the typical configurations for the investigation of QED cascades is the standing wave formed by two counter-propagating circularly polarized laser pulses with probe seeds located initially inside the wave~\cite{Kirk2009,PRSTAB054401,mercuribaron2024growth}.
The laser field plays a dual role by relativistically accelerating charged particles, \emph{i.e.}, electrons and positrons, to emit high-energy photons, and by interacting with the emitted photons to create electron-positron pairs which would be accelerated again to emit photons.
The growth of QED cascades depends on the field intensity and sensitively on the initial position of the probe seeds.
Around the magnetic nodes where the magnetic field is zero and charged particles are violently accelerated by the rotating electric field,
the numbers of the emitted photons and created pairs grow exponentially. 
The kinetic of charged particles is controlled by the Lorentz force and radiation recoil.
As the average effect of the Lorentz force, charged particles are expelled from the magnetic nodes and trapped finally to the electric nodes where the ponderomotive potential is minima and the photon emission and subsequent pair creation are suppressed.
The development of QED cascades is thus determined by the number of electron-positron pairs created by the emitted hard photons from one seed charged particle before it is trapped at an electric node.

The two elementary QED processes concerned in cascades are the nonlinear Compton scattering~\cite{PhysRevA705,PhysRevA022809,MRE0196125}, describing photon emission from a high-energy charged particle moving in strong electromagnetic fields, and nonlinear Breit-Wheeler process~\cite{breit34,Tang:2022a}, which describes the creation of an electron-positron pair in the collision of a high-energy photon with an intense laser pulse.
The outgoing particles in each process are generated mainly in a forward cone along the momentum of the incoming particle within a finite angular spread $\Delta \theta\propto 1/\gamma_{i}$~\cite{PRA2020012505,PRD056025},
which could lead to stochastic deflection of the outgoing particles' momentum from the direction of incoming particle's momentum, where $\gamma_{i}$ is the incoming particle's Lorentz factor.
This finite beaming effect of the outgoing particle's momentum has been considered to be negligible due to the ultra-relativistic particle energy $\gamma_{i}\propto\xi \gg1$ in laser-driven QED cascades,
and the outgoing particle is assumed to move with the collinear momentum of the incoming particle~\cite{Kirk2009,PRSTAB054401,mercuribaron2024growth}, where $\xi=|e|E_{l}/mc\omega_{l}\sim\mathcal{O}(10^{3})$ is the classical intensity parameter, $E_{l}$ ($\omega_{l}$) is the intensity (frequency) of the laser field, and $|e|$ ($m$) is the charge (mass) of the positron, $c$ is the speed of light.
However, as there is a huge number of photon emissions and pair creations in QED cascades, the finite beaming effect in each QED event could be accumulated to effectively change the momenta of later generated particles, and thus affect the long-term development of QED cascades.

In this paper, we investigate the importance of the finite beaming effect of outgoing particles' momenta on the growth of cascades by modeling QED events in simulations with the probability rate that is triple-differential in the particle energy as well as its transverse momentum.
We show that the beaming effect is negligible at the beginning of the QED cascades, but could considerably suppress the long-term development of the cascades. 
The rest paper is organised as follows.
We first outline the general method in Sec.~\ref{Sec_Theo}, modelling the elementary QED processes with the energy-and angularly resolved probability rate, benchmarking against the exact QED prediction,
and then apply the numerical model in our particle-tracking code in Sec.~\ref{Sec_simu} to simulate the QED cascades in circularly polarized standing waves and at the end conclude in~\ref{sec_con}.

\section{Simulation method}~\label{Sec_Theo}
The exact QED descriptions for the nonlinear Compton scattering and nonlinear Breit-Wheeler pair creation can be obtained from the S-matrix elements~\cite{Tang:2022a,MRE0196125}.
Here, we give the probability rate for the photon emission and pair creation based on the locally constant field approximation, as their formation length is much smaller than the scale of the field variation,~\emph{i.e.}, laser wavelength in our cases, at ultra-relativistic intensities $\xi\gg1$.
In our following discussion, we use the natural units $\hbar=c=1$, 
and introduce the dimensionless time $\omega_{l}t\to t$ and coordinates $\omega_{l}(x,y,z)/c\to (x,y,z)$, corresponding to the dimensionless period $T_{l}=2\pi$ and wavelength $\lambda_{l}=2\pi$ for the laser field.

The probability rate $W_{e}=\ud \trm{P}_{e}/\ud t$ for the photon emission by a high-energy charged particle with the Lorentz factor $\gamma_{p}$ can be written in the form~\cite{MRE0196125}
\begin{subequations}
\begin{align}
\frac{\ud W_{e}}{\ud \vepsilon_{e}}& =\frac{\alpha m}{\pi\gamma_{p}\omega_{l}} \iint \frac{\ud^{2}\bm{\ell}_{\LCperp}}{\vepsilon_{e}^{2}m^{2}}
\left[\eta_{p} \frac{1+(1-\vepsilon_{e})^{2}}{1-\vepsilon_{e}} - \mu\right] \mathrm{Ai}(\eta_{p})\label{Eq_NLC_Angle} \\
& =-\frac{\alpha m}{\gamma_{p}\omega_{l}}  \left[  \frac{1+(1-\vepsilon_{e})^{2}}{\mu(1-\vepsilon_{e})} \mathrm{Ai}^{\prime}(\mu)  + \trm{Ai}_{1}(\mu)\right]\,,\label{Eq_NLC_noAngle}
\end{align}
\label{Eq_NLC}
\end{subequations}
where $\alpha\approx1/137$ is the fine structure constant, $\vepsilon_{e}=\gamma_{\ell}/\gamma_{p}$ is the fraction of energy taken by the emitted photon from the incoming particle, $\bm{\ell}_{\LCperp}$ denotes the photon momentum in the plane perpendicular to the incoming particle's momentum $p^{\nu}$,
and $\eta_{p}=\mu[1+\ell^{2}_{\LCperp}/(\vepsilon_{e}m)^{2}]$, $\mu=\vepsilon_{e}^{2/3}/[(1-\vepsilon_{e})\chi_{p}]^{2/3}$,
$\chi_{p}=e|F^{\mu \nu} p_{\nu}|/m^{3}$ is the quantum parameter for photon emission, and $F^{\mu\nu}$ is the field-strength tensor.
$\trm{Ai}(x)$ and $\trm{Ai}^{\prime}(x)$ are the Airy function and its derivative, and $\trm{Ai}_{1}(x)=\int^{\infty}_{x} \trm{Ai}(y)\ud y$.

The probability rate $W_{b} = \ud\trm{P}_{b}/\ud t$ for the pair creation by a high-energy photon with energy $\gamma_{\ell} m$ is given as~\cite{Tang:2022a}
\begin{subequations}
\begin{align}
\frac{\ud W_{b}}{\ud \vepsilon_{b}}& =\frac{\alpha m}{\pi \gamma_{\ell}\omega_{l}} \iint \frac{\ud^{2} \bm{q}_{\LCperp}}{m^{2}}\left[\nu + \eta_{\ell} \frac{\vepsilon_{b}^{2}+(1-\vepsilon_{b})^{2}}{\vepsilon_{b}(1-\vepsilon_{b})}\right]\trm{Ai}(\eta_{\ell}) \label{Eq_NBW_Angle}\\
& =\frac{\alpha m}{\gamma_{\ell}\omega_{l}}  \left[\trm{Ai}_{1}(\nu)- \frac{\vepsilon_{b}^{2}+(1-\vepsilon_{b})^{2}}{\vepsilon_{b}(1-\vepsilon_{b})\nu} \trm{Ai}^{\prime}(\nu) \right]\,,\label{Eq_NBW_noAngle}
\end{align}
\label{Eq_NBW}
\end{subequations}
where $\vepsilon_{b}=\gamma_{q}/\gamma_{\ell}$ is the energy fraction taken by the created positron from the incoming photon,
$\bm{q}^{\LCperp}$ is the positron momentum in the plane perpendicular to the photon momentum $\ell^{\mu}$, and $\eta_{\ell}=\nu(1+q^{2}_{\LCperp}/m^{2})$, $\nu=[\vepsilon_{b}(1-\vepsilon_{b})\chi_{\ell}]^{-2/3}$,
$\chi_{\ell}=e|F^{\mu \nu} \ell_{\nu}|/m^{3}$ is the quantum parameter determining the creation process.

\begin{figure}[t!!!]
	\includegraphics[width=0.49\textwidth]{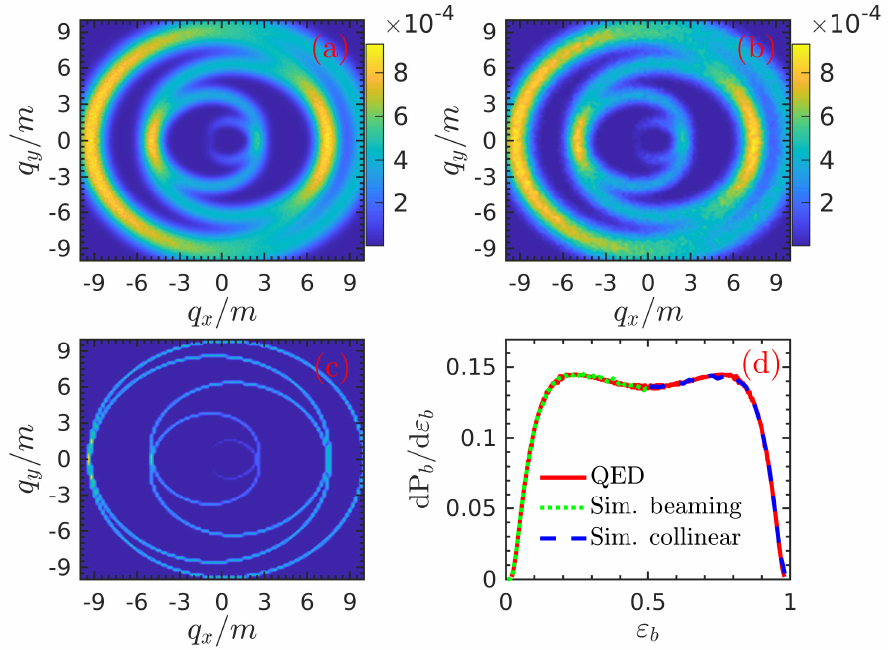}
	\caption {Comparison between the exact QED result (a) and simulation that does (b) and does not (c) include the finite beaming effect in pair creation: the transverse momentum profile of the created positron by a high-energy photon, $\gamma_{\ell}=10^{5}$, in the head-on collision with a circularly polarized laser pulse with the intensity $\xi=10$. The energy spectra of the created positron are compared in (d) between the QED result ($0<\varepsilon_{b}<1$), simulations with the finite beaming effect ($0<\varepsilon_{b}<0.5$) and with the collinear pair creation ($0.5<\varepsilon_{b}<1$) in terms of the symmetry at $\varepsilon_{b}=0.5$. The detail parameters are presented in the main text.}
	\label{Fig_QED_LCFA_sim}
\end{figure}

From Eqs.~(\ref{Eq_NLC_Angle}) and~(\ref{Eq_NBW_Angle}), we can reveal respectively the triple-differential spectra of the emitted photon and created positron, resolving the particle energy as well as its transverse momentum perpendicular to the momentum of the incoming particle, \emph{i.e.} the scattering angle.
Integrating the transverse momenta can then arrive at the particles' energy spectra [Eqs.~(\ref{Eq_NLC_noAngle}) and~(\ref{Eq_NBW_noAngle})],
which have been employed in the standard  particle-tracking code to sample the energy of the generated particles since the photon emission and pair creation are assumed to be collinear~\cite{Kirk2009,PRSTAB054401,mercuribaron2024growth}.
As an alternative, we implement the Monte-Carlo algorithm to sample the triple-differential spectra [Eqs.~(\ref{Eq_NLC_Angle}) and~(\ref{Eq_NBW_Angle})] in our simulations.
There needs three random numbers to determine the momentum of the outgoing particle in each QED event:
one is used to fix the energy fraction via the spectrum [Eqs.~(\ref{Eq_NLC_noAngle}) and~(\ref{Eq_NBW_noAngle})];
one is given to determine the modulus of transverse momentum via Eqs.~(\ref{Eq_NLC_Angle}) and~(\ref{Eq_NBW_Angle}) for the fixed energy fraction;
the last one is used as the random azimuthal angle along the momentum of the incoming particle.
The discrete photon emissions occur stochastically along a charged particle's trajectory, which is determined by the Lorentz force alone between two consecutive emissions, and the electron-positron pair is created in the annihilation of a high-energy photon with the probability rate calculated along the photon's classical trajectory.
The momentum of the recoiled charged particle after the photon emission and the paired electron in the creation is given by the conservation of momentum.


\begin{figure}[t!!!]
	\includegraphics[width=0.49\textwidth]{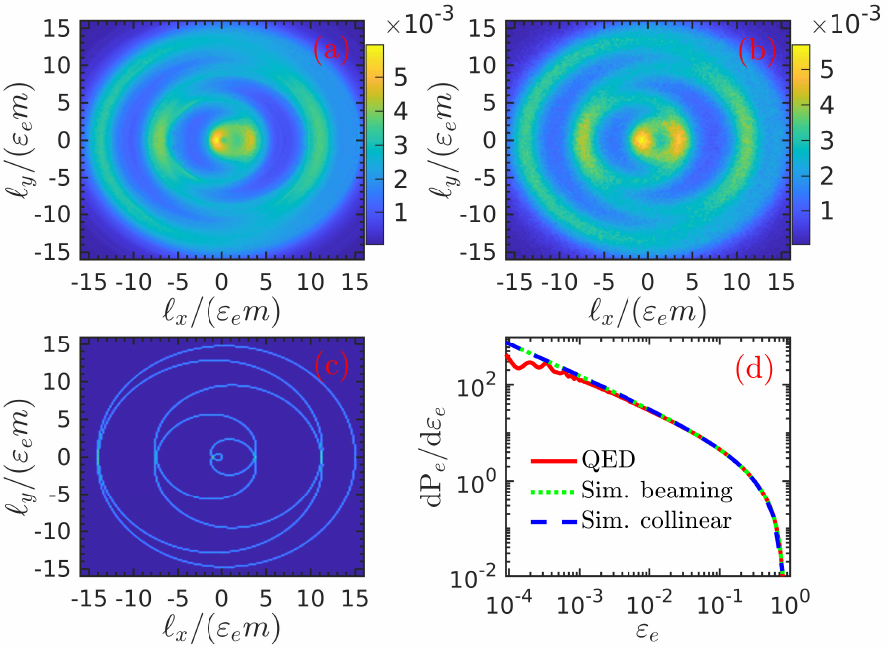}
	\caption {Comparison between the exact QED result (a) and simulation that does (b) and does not (c) include the finite beaming effect in the photon emission: the transverse momentum profile of the emitted photon by a high-energy electron, $\gamma_{p}=6000$, in the head-on collision with a circularly polarized laser pulse with the intensity $\xi=15$. The energy spectra of the emitted photon are compared in (d) between the QED result and simulations with and without the finite beaming effect. The other parameters are same as those in Fig.~\ref{Fig_QED_LCFA_sim}.}
	\label{Fig_QED_LCFA_sim_NLC}
\end{figure}

In Fig.~\ref{Fig_QED_LCFA_sim}, we benchmark our simulations against the exact QED calculation for pair creation, in which we consider the head-on collision between a high-energy photon with $\gamma_{\ell}=10^{5}$ and a circularly polarized laser pulse with the wavevector $k^{\mu}=\omega_{l}(1,0,0,-1)$ and vector potential $a^{\mu}(\phi) = m\xi(0,\cos\phi,\sin\phi,0)f(\phi)$, where $\omega_{l}=1.55~\trm{eV}$, $\xi=10$, $\phi$ is the laser phase, and the pulse envelope is $f(\phi)=\cos^{2}[\phi/(2N)]$ if $-N\pi<\phi<N\pi$ and $f(\phi)=0$ otherwise, $N=6$.
The photon annihilation after the pair creation and the subsequent photon emissions from the created pair are ignored.
The exact QED results in Fig.~\ref{Fig_QED_LCFA_sim} are calculated in the framework of Furry picture in the presence of a background laser field~\cite{Tang:2022a}.
As shown, accounting the finite beaming in the created positron momentum can improve substantially the agreement between the simulation and exact QED results: the transverse momentum profile predicted by the exact QED calculation in Fig.~\ref{Fig_QED_LCFA_sim} (a) can be well reproduced by the simulation in Fig.~\ref{Fig_QED_LCFA_sim} (b) when the finite beaming spread is included,
but collapses onto a curve corresponding the local vector potential along the photon trajectory if the finite beaming is ignored and the pair creation is assumed to be collinear in Fig.~\ref{Fig_QED_LCFA_sim} (c).
Both simulations with and without beaming effect can reproduce exactly the positrons' energy spectrum predicted by the QED result in \mbox{Fig.~\ref{Fig_QED_LCFA_sim} (d)}.

The comparison between the exact QED calculation~\cite{MRE0196125} and simulations with and without the finite beaming effect for the photon emission is presented in Fig.~\ref{Fig_QED_LCFA_sim_NLC} for $\gamma_{p}=6000$, $\xi=15$ and the same other parameters as in Fig.~\ref{Fig_QED_LCFA_sim}.
The radiation recoil to the incoming electron and the pair creation from the emitted photons are excluded in the simulations.
Similar as the discussion for pair creation, ignoring the beaming effect in the simulation would lose the accuracy in the prediction of the photons' transverse momentum profile as shown in Figs.~\ref{Fig_QED_LCFA_sim_NLC} (a)-(c)~\cite{PRA2020012505}, and can only reproduce the energy spectrum of the emitted photons in Fig.~\ref{Fig_QED_LCFA_sim_NLC} (d).
The infrared divergence of the LCFA spectrum~(\ref{Eq_NLC_noAngle}) results in the difference between the full QED result and simulations in the low-energy spectra $\vepsilon_{e}\to0$~\cite{king19a,PiazzaPRA2019}.
The parameters $\ell_{x,y}/(\vepsilon_{e}m)$ used as the independent variables in Figs.~\ref{Fig_QED_LCFA_sim_NLC} (a)-(c) correspond directly to the photon's scattering angles along the momentum of the incoming electron~\cite{MRE0196125}. 

\section{Simulation results}~\label{Sec_simu}
With our developed energy- and angularly resolved Monte-Carlo simulation method,
we intend to manifest the importance of the beaming effect on QED cascades. 
We consider QED cascades in the typical configuration of standing wave formed by two counter-propagating co-rotating circularly polarized laser pulses:
\begin{align}
E_{\LCp}&=\xi(~~\cos\phi_{\LCp},~~\sin\phi_{\LCp},0)~g(\phi_{\LCp},x,y)\,,\nonumber\\
B_{\LCp}&=\xi( -\sin\phi_{\LCp},~~\cos\phi_{\LCp},0)~g(\phi_{\LCp},x,y)\,,\nonumber\\
E_{\LCm}&=\xi(~~\cos\phi_{\LCm},~~\sin\phi_{\LCm},0)~g(\phi_{\LCm},x,y)\,,\nonumber\\
B_{\LCm}&=\xi(~~\sin\phi_{\LCm}, -\cos\phi_{\LCm},0)~g(\phi_{\LCm},x,y)\,,\nonumber
\end{align}
where $\phi_{\pm}=t\mp z$, $E_{\LCpm}$, and $B_{\LCpm}$ denote respectively the phase, electric and magnetic field of the laser propagating in the $\pm z$ direction and $g(\phi_{\pm},x,y)$ is the laser envelope.
A QED cascade can be characterized with the number of the electron-positron pairs $N(t)$ and its exponential growth rate $\Gamma(t)$, which could be calculated as
\begin{align}
\Gamma(t)=\ln[N(t+\Delta t/2)/N(t-\Delta t/2)]/\Delta t
\end{align}
where $\Delta t= T_{l}/10$ is used in our following calculations.

\begin{figure}
	\includegraphics[width=0.48\textwidth]{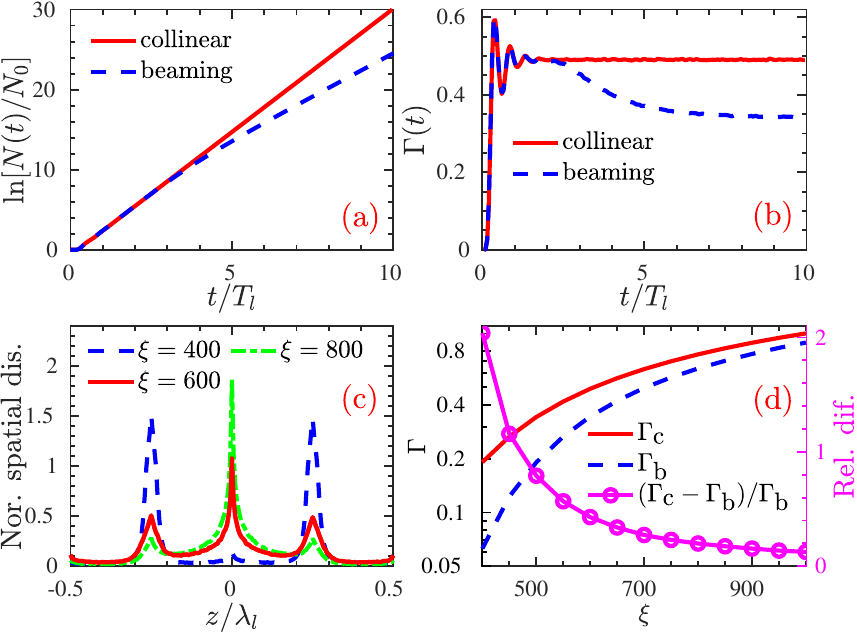}
	\caption {QED cascades with and without the beaming effect initialized by static seed pairs at the magnetic node in standing waves. (a) Number of electron-positron pairs and (b) its growth rate in the QED cascades for the laser intensity $\xi=600$.
(c) Normalized spatial distribution of the positrons when the QED cascades with the beaming effect get stabilized for different laser intensities $\xi=400,~600,~800$.  (d) Comparison of the pair number growth rate in stabilized QED cascades. The relative difference $(\Gamma_{c}-\Gamma_{b})/\Gamma_{c}$ (magenta circle line) increases from about $12.6\%$ at $\xi=1000$ to more than $200\%$ at $\xi=400$ shown with the right vertical axis.}
\label{Fig2_magnode}
\end{figure}

In Fig.~\ref{Fig2_magnode}, we first consider an ideal case of QED cascade initialized by $N_{0}=10^{5}$ pairs of electrons and positrons located statically at a magnetic node ($z=0$) of the standing wave formed by two monochromatic fields ($g=1$).
At the beginning of the cascade formation, 
the seed pairs are accelerated by the rotating electric field in the $x-y$ plane and get large values of energy and quantum parameter $\chi_{p}$.
As there is only a limited number of photon emissions and pair creations in this early stage,
the accumulation of the angular spread from each QED event could not affect the formation of the QED cascade as we can see in Figs.~\ref{Fig2_magnode} (a) and (b).
The kinetic of seed pairs are controlled dominantly by the lorentz force and causes the quantum parameter $\chi_{p}$ of the seed pairs oscillating
with the period of the field~\cite{PRSTAB054401}.
This oscillation can be transferred to that of emitted hard photons and leads to the oscillation of the growth rate of the electron-positron pairs in Fig.~\ref{Fig2_magnode} (b).

With the cascade proceeding, the number of the QED events increases exponentially and the narrow angular spread from each QED event could be accumulated to effectively change the momentum direction of later generated particles, and thus impact the development of QED cascades.
Without the beaming effect, the emitted hard photons and subsequently created electron-positron pairs are confined at the magnetic node ($z=0$) as their momenta are collinear in the $x-y$ plane.
This results in the exponential increase of the pair number $N(t)$ with a stable growth rate about $\Gamma_{c}=0.49$ after the formation of QED cascade, \emph{i.e.}, $t>2T_{l}$ for $\xi=600$, see the red solid line in Fig.~\ref{Fig2_magnode} (b).
As we can also see, when the finite beaming effect is included, the growth rate (blue dashed line) of the pair number decreases gradually and gets stabilized in a longer time at about $\Gamma_{b}=0.34$.
This is because the finite beaming effect in QED events deflects the momentum of the outgoing particle from the $x-y$ plane and thus releases the particles out of the magnetic node toward the region with lower electric field.
As shown in Fig.~\ref{Fig2_magnode} (c), in which the normalized spatial distribution of the positrons is plotted after the cascades including finite beaming effect get stabilized,
a large part of positrons are trapped at the electric node $z=\pm\lambda_{l}/4$ where the electric field is zero and the cascade is suppressed.

In Fig.~\ref{Fig2_magnode} (d), we compare the growth rate of the pair number with and without the beaming effect when QED cascades become stable.
As shown, the growth rates for both cases increase with the increase of the laser intensity~\cite{mercuribaron2024growth}, and ignoring the beaming effect in QED cascades would considerably overestimate the growth rate especially for lower intensities: the relative difference $(\Gamma_{c}-\Gamma_{b})/\Gamma_{b}$ increases from about $12.6\%$ at $\xi=1000$ to more than $200\%$ at $\xi=400$.
This tendency can be attributed to two reasons:
i) The higher intensity can accelerate particles to higher energy and thus results in narrower angular spread $\Delta\theta \propto 1/\gamma_{i}$ in QED events;
ii) As the probabilities for the photon emission~(\ref{Eq_NLC}) and pair creation~(\ref{Eq_NBW}) improve substantially for higher intensities,
electron-positron pairs could be created efficiently in a broader region and heaped up tightly around the magnetic node before a seed electron or positron is trapped around an electric node. This gives rise to the particle spatial distribution getting closer to the particle confine at the magnetic node without the beaming effect.
As shown in Fig.~\ref{Fig2_magnode} (c), a larger fraction of the positrons would distribute around the magnetic node at higher intensities, even the beaming effect is included.

\begin{figure}[t!!!]
	\includegraphics[width=0.49\textwidth]{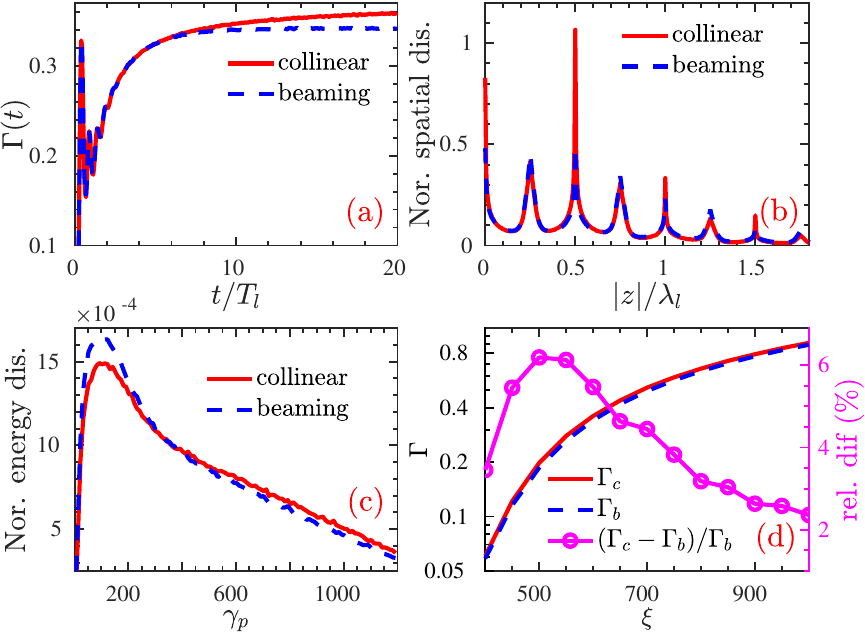}
	\caption {QED cascades initialized by seed pairs with an uniform distribution in the region of $|z|<\lambda_{l}/4$ and $|p_{z}|<500m$ with (blue dashed lines) and without (red solid lines) finite beaming effect.
(a) Growth rate $\Gamma(t)$ of the pair number in the cascades for the laser intensity $\xi=600$, and the corresponding normalized spatial (b) and energy (c) distribution of positrons at $t=20T_{l}$.
(d) Pair number growth rate with and without finite beaming effect at $t=20T_{l}$ for different intensities.
Their relative difference (magenta circle line) is shown with the right vertical axis.}
	\label{Fig3_unidist}
\end{figure}

In Fig.~\ref{Fig3_unidist}, we consider QED cascades initialized by seed electron-positron pairs ($N_{0}=10^{5}$) with an uniform distribution in the spatial $|z|<\lambda_{l}/4$ and momentum $|p_{z}|<500m$ region.
As shown in \mbox{Fig.~\ref{Fig3_unidist} (a)}, the finite beaming can hardly affect the early formation of the cascades, but considerably decreases the growth rate of the pair number in the later development of the cascades.
The growth rate without beaming effect is increasing in the long-term proceeding of the QED cascade as more and more particles are generated and tightly heaped up around the magnetic nodes $z=n\lambda_{l}/2$, where $n$ is an integer,
while the particle heaping up in this high-field region could be effectively balanced by the finite beaming effect as shown in \mbox{Fig.~\ref{Fig3_unidist} (b)},
in which we plot the normalized spatial distribution of the positrons at $t=20T_{l}$ and can clearly see that the fraction of positrons around the magnetic nodes is reduced by the finite beaming effect.
This reduction also results in the less acceleration of the charged particles in the cascade as shown in \mbox{Fig.~\ref{Fig3_unidist} (c)}.

We point out that when the QED cascade including the finite beaming effect gets stabilized, the normalized particle's spatial distribution in each laser cycle is almost unchanged and depends weakly on the initial conditions of the cascade, as the initial spatial (momentum) distribution of the seed pairs would be smoothed (averaged) by the generated particles.
This steady distribution would contribute to the same growth rate for a stable cascade, see the blue dashed line in \mbox{Fig.~\ref{Fig3_unidist} (a)} and compare with that in \mbox{Fig.~\ref{Fig2_magnode} (b)}.
For uniformly distributed seed pairs, the relative importance of the beaming effect on the growth rate is much weaker, $(\Gamma_{c}-\Gamma_{b})/\Gamma_{b}< 7\%$ in Fig.~\ref{Fig3_unidist} (d), than the case in Fig.~\ref{Fig2_magnode} (d),
as the Lorentz force can also work to expel the charged particles (not exactly at the magnetic node) out of the high-field region,
while for seed pairs at the magnetic node in Fig.~\ref{Fig2_magnode}, it is the beaming effect plays the role to release them out of the magnetic node.
Different from the tendency in \mbox{Fig.~\ref{Fig2_magnode} (d)},
the finite beaming effect is less important for lower intensities $\xi<500$ as shown in Fig.~\ref{Fig3_unidist} (d). This is because the photon emission and pair creation is inefficient and the reduction of the particle heaping up around the magnetic node cannot be effective.
With the substantial improvement of the emission and creation probabilities at higher intensities $\xi>600$,
the relative importance of the finite beaming effect also decreases as QED events could be efficient in a much broader region apart from the magnetic node. 
These two reasons contribute to the tendency of the relative difference between the growth rates in \mbox{Fig.~\ref{Fig3_unidist} (d)}.
We also point out that the slight difference in the growth rates could still lead to considerable difference in the pair yield of the long-term cascades.

We now move to a more realistic situation in which two circularly polarized laser pulses counter-propagate with the envelope
\[g(\phi,r) = \frac{1}{2}\left(\tanh\frac{\phi+\phi_{0}}{\sigma_{l}}-\tanh\frac{\phi-\phi_{0}}{\sigma_{l}}\right)~e^{-r^{2}/\sigma^{2}_{t}}\]
where $\sigma_{l} = 0.5T_{l}$ scales the rising and falling edges of the pulse, and $\sigma_{t}=5\lambda_{l}$ scales the transverse width of the pulse, the parameter $\phi_{0}$ controls the longitudinal duration, \emph{i.e.}, full width at half maximum $D=2\phi_{0}$, of the pulse.   
At the beginning of the simulation \mbox{$t=0$}, the separation between two laser fronts is about $3T_{l}$, and the seed pairs ($N_{0}=10^{5}$) distribute uniformly in the spatial region$|z|<\lambda_{l}/4$ with the initial momentum $|p_{z}|<500 m$.

\begin{figure}[t!!!]
	\includegraphics[width=0.48\textwidth]{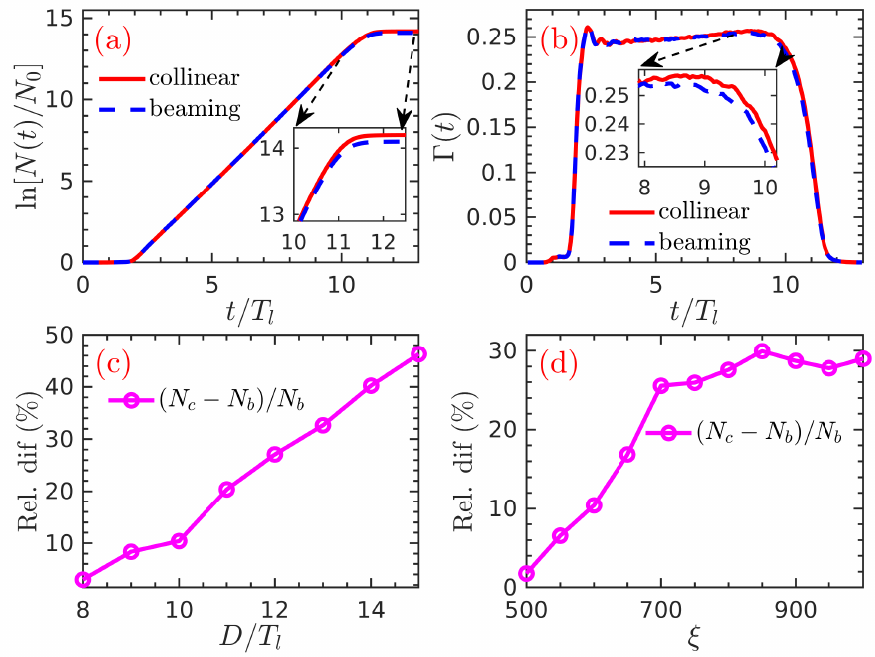}
	\caption {QED cascades driven by two counter-propagating laser pulses with circular poalrization.
 (a) Pair number and (b) its growth rate in the cascades with (blue dashed lines) and without (red solid lines) finite beaming effect for the laser intensity $\xi=600$ and pulse duration $D=10T_{l}$.
 The dependence of the relative difference between the pair yield with ($N_{b}$) and without ($N_{c}$) beaming effect on the pulse duration $D$ and intensity $\xi$ are given, respectively, in (c) for the fixed laser intensity $\xi=600$ and (d) for the pulse duration $D=10T_{l}$. }
	\label{Fig_twopulse}
\end{figure}

Fig.~\ref{Fig_twopulse} (a) shows the growth of the pair number during the QED cascade driven by the pulses with $\xi=600$ and $\phi_{0}=5T_{l}$.
The cascade could sustain for the duration about $D=10T_{l}$, induced by the seed pairs when the two laser pulses start to overlap around $t=1.5T_{l}$, and ceases at around $t=11.5T_{l}$ when the two laser pulses separate.
Fig.~\ref{Fig_twopulse} (b) presents the growth rate of the pair number during the cascade.
The smooth peak in the growth rate around $t=1.2 T_{l}$ comes from the pair creation in the collision between the high-energy seed pairs and the propagating laser pulses.
Again, the finite beaming effect cannot affect the early stage of the cascade development, but decreases the particle growth in the long-term cascades.
As shown in the inset in \mbox{Fig.~\ref{Fig_twopulse} (a)}, ignoring the finite beaming effect could considerably overestimates the pair yield of the QED cascade by about $(N_{c}-N_{b})/N_{b}=10.4\%$, even though the relative difference between the maximal growth rate with and without beaming effect is just about $(\Gamma_{c, \trm{max}}- \Gamma_{b, \trm{max}})/\Gamma_{b, \trm{max}}=1.0\%$, see the inset in \mbox{Fig.~\ref{Fig_twopulse} (b)}, where $N_{b}$ ($N_{c}$) is the number of the pairs at the end of the cascades with (without) beaming effect.

As we can see in Fig.~\ref{Fig_twopulse} (c) for $\xi=600$, the importance of the finite beaming effect on the pair yield depends linearly on the duration of the cascade with the relative difference increasing from about $3\%$ for $D=8T_{l}$ to about $46\%$ for $D=15\%$. This can be simply understood as
\begin{align}
\frac{N_{c}-N_{b}}{N_{b}} \propto e^{(\Gamma_{c}-\Gamma_{b}) D}-1 \approx (\Gamma_{c}-\Gamma_{b}) D
\end{align}
where $(\Gamma_{c}-\Gamma_{b}) D\ll1$ as shown in Fig.~\ref{Fig_twopulse} (b).
To further improve the importance of the beaming effect on the pair yield, we can also increase the pulse intensity as shown in Fig.~\ref{Fig_twopulse} (d) for the fixed pulse duration $D=10T_{l}$. The relative difference would increase to about $30\%$ for the laser intensity higher than $\xi=800$ because of the increase of the difference ($\Gamma_{c}-\Gamma_{b}$) between the growth rate.

\section{Conclusion}~\label{sec_con}

We develop a new particle-tracking code, which can resolve not only the energy spectra but also the beaming spread of the outgoing particles' momenta in QED events.
Accounting the finite beaming spread in the outgoing particle momentum can improve substantially the agreement between simulations and exact QED results.

We investigate the finite beaming effect on the growth of QED cascades driven by two counter-propagating co-rotating circularly polarized laser pulses.
The finite beaming effect can hardly affect the early formation of QED cascades, but can considerably reduce the growth of the particle number in the long-term development of cascades.
For the ideal cascades initialized by the static seed pairs at the magnetic nodes,
the finite beaming effect can significantly reduce the growth rate of the pair number by more than $200\%$ at $\xi=400$ and about $12.6\%$ at $\xi=1000$.
For the more realistic cascades in the collision of two finite laser pulses, the finite beaming effect can only induce slight difference in the particle number growth rate, but can lead to the considerable decrease of the particle yield especially at ultrahigh intensities. 
For QED cascades driven by the pulses longer than $10$ laser cycles with the intensity $\xi>600$, the finite beaming effect could decrease the final pair yield by more than $10\%$.

In our discussion, we manifest that the finite beaming affects QED cascades in the way by reducing the particle heaping up in the region with high electric field.
Therefore, we can infer that the finite beaming effect may not influence significantly the cascades in the field configuration without stable high-field points, such as the standing wave formed by two counter-propagating linearly polarized laser pulses and that by circularly polarized laser pulses with counter rotation.
As the effect of the finite beaming can only be brought out in the long-term developed QED cascades, its importance should be further checked with the more self-consistent codes including the laser energy absorption~\cite{PRL035001} and the electron-positron plasma effects~\cite{gonoskov15}.

\section{Acknowledgments}
The author acknowledges support from the Shandong Provincial Natural Science Foundation, Grants No. ZR202102280476.
The work was carried out at Marine Big Data Center of Institute for Advanced Ocean Study of Ocean University of China.

\bibliographystyle{apsrev}
\providecommand{\noopsort}[1]{}

\end{document}